\documentstyle[12pt]{article}
\begin{document}

\title{{\Large \bf NEUTRINO MASS} \footnote{To be published in the {\em
Proceedings of
the  European Physical Society Meeting}, HEP93, Marseille, July 1993. }}
\author{\bf Goran Senjanovi\'c\\
        \it International Centre for Theoretical Physics\\
        \it Trieste, Italy \\}

\maketitle

\begin{abstract}
We review the issue of neutrino mass by concentrating on the minimal
 extensions of the standard model. In particular, we emphasize the
 role that gravitation may  play in this regard and discuss the
 central aspects of the see-saw mechanism of generating neutrino mass,
 including the possibility of the see-saw scale being close to the
 electroweak scale.
\end{abstract}

\newpage

The experimental limits on neutrino masses are [1]:
\begin{equation}
m(\nu_{e}) \leq   8 eV  ,\; \;  m(\nu_{\mu}) \leq 220 keV ,
 \; \; m(\nu_{\tau}) \leq  31 MeV
\end{equation}

Why are neutrinos much lighter than the charged leptons? If they are
not massless, can the theory shed any light on the values of their
 masses? Can neutrino mass be responsible for the solar neutrino
puzzle and (or) dark matter of the universe? We summarize briefly
 the situation regarding the above questions keeping the theoretical
 framework as simple as possible. \vskip 4.0mm
{\bf \noindent I. The standard model and the neutrino mass. }
\vskip 2.0mm
It is a text-book knowledge that $m_{\nu}=0$ in the standard model.
 Is it necesarily so? Define first the standard model as a theory
 without $\nu_R$ (as is commonly done). The only mass term for
 neutrino is necessarily of Majorana type $\nu_L^TC\nu_L$ and
 this is forbidden by the B-L accidental U(1) symmetry of the
 standard model. Thus, neutrino is massless, to all orders in
 perturbation theory. Furthermore, since B-L is anomaly free
 there can be no nonperturbative weak interaction effects that
 could induce a tiny $m_{\nu}$. There seems to be no possible
loophole in the argument.

Now, what if gravity breaks B-L symmetry? We know that global
 symmetries may have no meaning in the vicinity of black holes.
 If so, we have no right to assume that the effective theory
 respects this symmetry (however, $SU(2)\times U(1)$, being a
local gauge symmetry, would still remain unbroken). The relevant
leading order operator that breaks B-L has the form [2]

\begin{equation}
\Delta L_{\rm eff} \simeq c_{ij}(\ell_{i}^T C \tau_2 \vec{\tau}
\ell_j) {(\phi^T  \tau_2 \vec{\tau} \phi) \over M_{pl} }
\end{equation}

where $\ell_i = \left(\begin{array}{c} \nu_i \\ e_i \end{array}
\right)_L$  and $\phi $ is the usual Higgs doublet; and $M_{pl}$
is the Planck scale  and $c_{ij}$ are numbers (of order one?).
This should imply a neutrino mass matrix

\begin{equation}
(M)_{ij} = c_{ij} {<\phi>^2 \over M_{pl}}  \equiv c_{ij} m_{\nu}
\end{equation}

where $m_{\nu}$ sets the neutrino mass scale, $m_{\nu}  \simeq 10^{-5} eV$.
This appears ridiculously small and at first glance of academic interest
 only. What about $c_{ij}$?
One possibility is $c_{ij}=c \delta_{ij}$, in which case all $\nu$'s
 would be degenerate: $\Delta m_{\nu}^2=0$. But, on the other hand,
 we could imagine gravity leading to a ``democratic'' mass matrix;
 $c_{ij}=1$ for all $i, j = 1,...,N$. For $N=2$, this
would imply
\begin{equation}
M_{\nu} = m_{\nu}\left(\begin{array}{cc} 1&1\\1&1\end{array}\right)
\end{equation}

or $m_{\nu}^{(1)} = 2 m_{\nu}$, $m_{\nu}^{(2)} = 0$ and
 $\Delta m_{\nu}^2 = 4 m_{\nu}^2 \simeq 10^{-10} eV^2$,
$\theta_{\rm mix} = 45^o$.

The mass difference $\Delta m_{\nu}^2  \simeq 10^{-10} eV^2$ with
 the maximal mixing is precisely what is needed for the so called
``just so'' long oscillation length solution of the  solar
 neutrino puzzle (SNP). The electron neutrino leaves the sun and
 by the time it arrives to Earth it is turned into  the muon neutrino.
 Similar conclusion is reached for the realistic case $N=3$.

In short, contrary to the conventional wisdom, the standard model as
 it stands may offer a resolution to the SNP (as long as you are
willing to accept gravity as part of the theory). On the other hand,
 no other interesting consequence for the neutrino physics:
 no dark matter and no neutrino-less double  $\beta$ decay.

\vskip 4.0mm
{\bf \noindent II. The see-saw mechanism.}
\vskip 2.0mm

Things can change drastically if you add a righthanded neutrino to
the model (a minimal change).
The $SU(2) \times U(1)$ gauge symmetry implies the well-known form of
 the neutrino mass matrix [3]

\begin{equation}
\begin{array}{c} \nu_L\\
\nu_R  \end{array} \left( \begin{array}{cc}  0 & m_D \\
 m_D & M_R \end{array} \right)
\end{equation}

where you expect an $SU(2) \times U(1)$ invariant mass for $\nu_R$,
 $M_R \gg m_D$ (Dirac mass term). This implies the celebrated see-saw
 induced small mass for $\nu_L$
\begin{equation}
m_{\nu} \simeq  {m_D^2 \over M_R}
\label{emenu}
\end{equation}

The neutrino is light not because $m_D \ll  m_e$, but since its
righthanded counterpart is very heavy: $M_R \rightarrow \infty$
 means $m_{\nu} \rightarrow  0$.
Of course, in order to predict $m_{\nu}$ we must know where $M_R$
 lies (and also somehow determine $m_D$). Let us discuss some
 possible values of  $M_R$ associated  with different ideas about
 its origin.

\vskip 4.0mm
{\it \noindent A. Planck scale as $M_R$ (gravity again).}
\vskip 2.0mm
Certainly a natural value, after all this is the only new physics
 scale we know of. This means we are back in the situation
 described in {\bf I}, when gravity is assumed to break a global
 B-L symmetry.

\vskip 4.0mm
{\it \noindent B. $M_R$  and left-right symmetry. }
\vskip 2.0mm

In $SU(2)_L \times SU(2)_R \times U(1)_{B-L}$ left-right symmetric
 theories $M_R$ is the scale of $SU(2)_R$ and B-L (gauged) symmetry
 breaking. The breaking is achieved with the Higgs fields $\Delta_L$
and $\Delta_R$, triplets under $SU(2)_L$ and $SU(2)_R
$ respectively, carrying two units of B-L, so that $M_R \simeq
<\Delta_R>$ [4].

What about $M_R$? Experiment tells us only $M_R > 1 TeV$. Here
 I just wish to report on the interesting possibility of $M_R
\simeq 1 TeV$;
if you also assume $m_D \simeq m_e$ (charged lepton mass), you
end up with a 
spectrum of neutrino masses

\begin{equation}
m(\nu_{\tau}) \simeq 1 MeV , \; \; m(\nu_{\mu}) \simeq 10 keV ,
 \; \; m(\nu_{e}) \simeq 1 eV
\end{equation}

interestingly close to the experimental limits.
This forces $\nu_{\tau}$ and $\nu_{\mu}$ to decay (it is
 certainly possible) [5] and $\nu_{e}$ could be  the dark matter
 of the universe.
Furthermore, we could {\em expect visible neutrino-less double
 $\beta$ decay} [4], but no impact on the SNP.

Of course, nothing prevents us from taking $M_R$ to be large.
In this case, it makes sense to embed $SU(2)_L \times SU(2)_R
\times U(1)_{B-L}$ into  a $SO(10)$ grand unified theory which
can give us a hint on $M_R$ and also relate $m_D$ to $m_L$.

\vskip 4.0mm
{\it \noindent C. $M_R$  and grand unification.}
\vskip 2.0mm
In $SO(10)$ $\nu_R$ completes the one-family content of a 16-dim.
 spinorial representation and $\Delta_L$ and $\Delta_R$ belong
 to a 126 dimensional Higgs.  In the minimal version one
 predicts $ m_D = m_u$ and thus from (\ref{emenu})
  $m(\nu_{\tau}):m(\nu_{\mu}):m(\nu_{e})= m_t^2:m_c^2:m_u^2$ under
 the assumption that $M_R$ is generation blind (?). In that case
 you have an interesting possibility of $ m(\nu_{\tau}) \simeq
 (1 -10) eV$ needed for dark matter, $m(\nu_{\mu}) \simeq (10^{-4}
 -10^{-3}) eV$ and
$m(\nu_{e}) \simeq (10^{-9} -10^{-8}) eV$, allowing for an  MSW
 solution [1] of the SNP. Also,
the mixing angles may lie in the desired range, but I will not
 discuss it here due to  lack of space. The above possibility
is realized with  $M_R$ being an intermediate symmetry
 breaking mass scale.

However, even the qualitative features of the predictions depend
 on the nature of symmetry breaking. If instead of  $126_H$
you use $16_H$ to break $SU(2)_R$ and B-L, $M_R$ gets induced
at the two-loop level  [6] and $M_R \propto m_u$, which in
 the minimal theory gives $m_{\nu} \propto m_u$ (and not
 $m_u^2$). For $ m(\nu_{\tau}) \simeq(1 -10) eV$, now
$m(\nu_{\mu}) \simeq (10^{-2} -10^{-1}) eV$ and
$m(\nu_{e}) \simeq (10^{-5} -10^{-3}) eV$, which would be
 outside the MSW range.

\vskip 4.0mm
{\it \noindent D.  Majorons: $M_R$ and global B-L breaking.  }
\vskip 2.0mm

Of course, B-L may not be gauged and its spontaneous breaking
would then result in a Goldstone boson, the Majoron [7].
 The result is still the see-saw as in (6), only $M_R =
 <\sigma>$, where $\sigma$ is an $SU(2)\times U(1)$
 singlet field with $(B-L) \sigma = 2 \sigma$.
Again, $M_R$ = ?

An interesting possibility materializes if one takes seriously
 the gravitationally induced breaking of global symmetries
 [8], [9].  The Majoron gets a mass  and has to decay into
 $\nu \nu $ pairs in order not to upset the standard
 cosmological scenario.
Now, the $J\nu\nu$ coupling is of order $<\sigma>^{-1}$, {\em
 hence an upper limit on $<\sigma> < 1 TeV$}, which implies a
 {\em lower limit on the heaviest neutrino mass, $m_{\nu}\geq
 (0.1-1)eV $} [9]. In other words, it is expected that  the
neutrino makes some portion of the dark matter of the universe.

\vskip 4.0mm
{\bf \noindent III. Summary.}
\vskip 2.0mm

As I have tried to demonstrate, the theory of neutrino mass
suffers from a grave setback, an inability to  provide a
 prediction. It is highly suggestive that the neutrino is
a Majorana particle, for then the see-saw mechanism offers
a natural explanation
of the smallness of $m_{\nu}$. The trouble is that $m_{\nu}$
depends sensitively on $M_R$, the mass of the right-handed
neutrino and  $M_R$ may lie anyway between $ 1 TeV $ and
 $10^{19} GeV$ [10]. The end points are kind of interesting.
 For $M_R \simeq 1
TeV$, you end up with neutrino mass close to the
experimental limits and you would expect $\nu_e$
 to provide a dark matter and also to give observable
 neutrino-less double $\beta$ decay. The Planck scale
value, $M_R \simeq 10^{19} GeV$, on the other hand
, could gives us naturally the ``just so'' oscillation
 solution to the SNP just by including gravity to the
standard model.

The values for $M_R$ in between, not surprisingly, offer
a large range of different possibilities for $m_{\nu}$.
 You may  have $\nu_{\tau}$ as dark matter and you may
 have MSW solution to the SNP and you may even have them
 simultaneously in the attractive
 case of $M_R$ being an intermediate scale in $SO(10)$.
 It is clear that we need more experimental information
 desperately.

\vskip 4.0mm
{\bf \noindent Acknowledgements}
\vskip 2.0mm

I  thank A. Smirnov for  useful discussions

\vskip 4.0mm
{\bf \noindent References and notes}
\vskip 2.0mm

\noindent
[1]  See, e.g. A. Smirnov,  ICTP preprint IC/93/359 for
 a recent review. A nice treatise on neutrino physics
is R.N. Mohapatra and P. Pal, World Scientific (1991).
\newline

\noindent
[2] R. Barbieri, J. Ellis and M.K. Gaillard,
{\it Phys. Lett.} {\bf B90}
(1980)  249; E. Akhmedov, Z. Berezhiani and G. Senjanovi\'c,
{\it Phys. Rev. Lett.} {\bf 69}
(1992) 3013.
\newline

\noindent
[3] M. Gell-Mann, P. Ramond and R. Slansky, in {\it Supergravity},
 ed. by P. van Niewenhuizen and D. Freedman (Amsterdam,
 North Holland, 1979) 315; T. Yanagida, in Proc. of the
Workshop on the {\em Unified Theory and Barion Number in
the Universe}, eds. O. Sawada and A. Sugamoto
 (KEK, Tsukuba) 95 (1979);  R. N.
Mohapatra and G. Senjanovi\'c,
{\it Phys. Rev. Lett. } {\bf 44}
(1980) 912.
\newline

\noindent
[4] Mohapatra and Senjanovi\'c, in [3]. The problem of induced
 vev $<\Delta_L> \neq 0$ can be solved by breaking parity above
the scale $M_R$, see D. Chang and R.N. Mohapatra,
 {\it Phys. Rev.  } {\bf D32}
(1985) 1248.
\newline

\noindent
[5] M. Roncadelli and G. Senjanovi\'c,
{\it Phys. Lett.} {\bf B107}
(1983) 59; also P. Herzog and R.N. Mohapatra, {\it Phys.
 Rev. Lett. } {\bf 67 }
(1992) 2475.
\newline

\noindent
[6] E. Witten ,
{\it Phys. Lett.} {\bf B91}
(1980) 81.
\newline

\noindent
[7] Y. Chikashige, R. N. Mohapatra and R.D. Peccei,
{\it Phys. Lett.} {\bf B98}
(1981) 265.
\newline

\noindent
[8] D. Grasso, M. Lusignoli and M. Roncadelli ,
{\it Phys. Lett.} {\bf B288 }
(1993) 140 .
\newline

\noindent
[9] A. Akhmedov, Z. Berezhiani, R. N. Mohapatra and G.
 Senjanovi\'c,
{\it Phys. Lett.} {\bf B299}
(1993) 90.
\newline

\noindent
[10] It is even possible that $\nu_R$ may escape from getting
 heavy and remain as a light sterile particle. This and other
 issues, such as the radiative generations of neutrino mass
 are omitted only due to the lack of space. For a recent work,
see for example J.T. Peltoniemi and J.W.F. Valle, {\it Nucl. Phys.}
 {\bf B406} (1993) 409, and references therein.

\end{document}